# LOCALLY D-OPTIMAL DESIGNS BASED ON A CLASS OF COMPOSED MODELS RESULTED FROM BLENDING EMAX AND ONE-COMPARTMENT MODELS[1]

By X. Fang and A. S. Hedayat

*University of Illinois, Chicago*

A class of nonlinear models combining a pharmacokinetic compartmental model and a pharmacodynamic Emax model is introduced. The locally D-optimal (LD) design for a four-parameter composed model is found to be a saturated four-point uniform LD design with the two boundary points of the design space in the LD design support. For a five-parameter composed model, a sufficient condition for the LD design to require the minimum number of sampling time points is derived. Robust LD designs are also investigated for both models. It is found that an LD design with $k$ parameters is equivalent to an LD design with $k-1$ parameters if the linear parameter in the two composed models is a nuisance parameter. Assorted examples of LD designs are presented.

**1. Introduction.** A class of models is constructed by plugging a pharmacokinetic (PK) compartmental model into a pharmacodynamic (PD) Emax model. Under this class of models, only one measurement is required per study subject rather than multiple measurements and both PK and PD parameters can be estimated by a single experimental setup. Other advantages of this approach will be listed shortly after the related PK and PD models are introduced.

A basic PD Emax model can be expressed as $E([D]) = E_0 + E_{\max}[D]/(ED_{50} + [D])$, where $E(\cdot)$ is the drug effect such as reduction in low-density

Received April 2006; revised March 2007.
[1]Supported by NSF Grants DMS-01-03727, DMS-06-03761 and NIH Grant P50-AT00155 (jointly supported by National Center for Complementary and Alternative Medicine, the Office of Dietary Supplements, the Office of Research on Women's Health and National Institute of General Medicine). Any opinions, findings, and conclusions or recommendations expressed in this material are those of the authors and do not necessarily reflect the views of the NSF and the NIH.

cholesterol and $[D]$ is the concentration of free drug in the environs of the drug receptor. The Emax model contains three PD parameters. $ED_{50}$ is the drug concentration showing 50% of the maximum drug effect, $E_{\max}$ is the maximum drug effect and $E_0$ is the baseline effect. For the concept of Emax model, the reader is referred to Ritschel [22]. Various applications of Emax models have been discussed by, but are not limited to, Graves et al. [7], Demana et al. [4], Angus et al. [1], Staab et al. [26] and Hedayat, Yan and Pezzuto [11, 12].

Since a PK compartmental model describes drug concentration across time, the $[D]$ can be approximated by an appropriate PK model. After replacing $[D]$ in the Emax model by an open one-compartment model with IV bolus input and first-order elimination results in the composed Emax-PK1 model

$$(1.1) \qquad Y_{tj} = \beta_0 + \frac{\beta_1 \cdot D}{\beta_2 e^{\beta_3 t} + D} + \varepsilon_{tj},$$

where $Y_{tj}$ is the effect of the drug at time $t$ observed on subject $j$, $\beta_0$ is the minimum residual effect, $\beta_1$ is the maximum drug effect, $\beta_2$ is the $ED_{50}$ and $\beta_3$ is the total elimination rate. It is assumed that the errors across time and subjects are i.i.d. $N(0, \sigma^2)$. Throughout this paper, $D$ is the administered dose in the unit of drug amount per unit body weight. For the concept and essential results of the one-compartment model related to IV bolus, the reader is referred to Rowland and Tozer [24], Landaw [16], Dette and Neugebauer [5], Han and Chaloner [8] and Hedayat, Zhong and Nie [10].

If $[D]$ is replaced by an open one-compartment model with the first-order input and the first-order elimination, the resulting composed Emax-PK2 model becomes

$$(1.2) \qquad Y_{tj} = \beta_0 + \frac{\beta_1 \cdot D(e^{-\beta_2 t} - e^{-\beta_3 t})}{\beta_4(1 - \beta_3 \beta_2^{-1}) + D(e^{-\beta_2 t} - e^{-\beta_3 t})} + \varepsilon_{tj},$$

where $\beta_0$ is the baseline effect, $\beta_1$ is $E_{\max}$, $\beta_2$ is the absorption rate, $\beta_3$ is the total elimination rate and $\beta_4$ is the $ED_{50}$. The assumption about the error terms is the same as that in model (1.1). For the concept and various results related to this one-compartment model, the reader is referred to Rodda, Sampson and Smith [23], Davidian and Gallant [3], Mallet [18], Mandema, Verotta and Sheiner [19], Verme et al. [27], Lindsey et al. [17], Landaw [16], Atkinson et al. [2] and Wakefield [28].

Other advantages of the Emax-PK models include (1) The PK parameters can be estimated without drawing blood samples. (2) Drug effect becomes a function of time rather than a function of drug concentration which itself is a function of time. As a result, the drug effect is predictable across time, such as blood pressure is being reduced across time. (3) Sampling at various time points is controllable before taking samples, whereas the drug concentration in an Emax model alone is unknown before sampling. Therefore, one can obtain better estimates of PK and PD parameters by implementing suitable designs. (4) The residual effect of a drug can be estimated at any time point through the Emax-PK1 model. (5) The baseline effect of a drug can be estimated through the Emax-PK2 model.

In this paper, locally D-optimal (LD) designs and robust LD designs for the preceding two Emax-PK models are studied for the purpose of estimating all population PK parameters. These PK parameters are considered fixed effects although they likely differ between subjects. In Section 2, the equivalence theory for nonlinear models is briefly reviewed. Section 3 contains the main results about the number of sampling time points in an LD design support for the Emax-PK models. Examples of LD designs for both models are provided in Section 4. Some assorted robust LD designs are investigated in Section 5. Discussion and conclusion are summarized in Section 6.

**2. Preliminaries.** The equivalence theorem for designs based on linear models was first introduced and developed by Kiefer and Wolfowiz [15]. White [29] extended the equivalence theorem to locally optimal designs based on nonlinear models. In a nonlinear model, the Fisher information matrix depends on unknown parameters. As a result, there is no global optimization for all values of model parameters and the optimality of a design can be evaluated only for special values of model parameters, called nominal or postulated values. For details, the reader is referred to Silvey [25] and Hedayat [13].

Without loss of generality, let us consider the nonlinear model $Y_{tj} = \eta(t,\vec{\beta}) + \varepsilon_{tj}$ with $\varepsilon_{tj}$'s being i.i.d. $N(0,\sigma^2)$. The normalized Fisher information matrix for the entire model parameters based on the design measure $\xi = \{(t_1,p_1),\ldots,(t_K,p_K)\}$ can be expressed as

$$(2.1) \qquad M(\xi,\vec{\beta}) = \sigma^{-2} \sum_{i=1}^{K} p_i \left(\frac{\partial \eta(t_i,\vec{\beta})}{\partial \vec{\beta}}\right)\left(\frac{\partial \eta(t_i,\vec{\beta})}{\partial \vec{\beta}}\right)'.$$

A design measure $\xi = \{(t_1,p_1),\ldots,(t_K,p_K)\}$ is a description of sampling time points $(t_1,\ldots,t_K)$ where $Y_{t_ij}$ will be measured for the $j$th subject at time $t_i$. Associated with $t_i$ is the mass $p_i$ such that $0 < p_i < 1$, and $\sum p_i = 1$. In design $\xi$, $p_1,\ldots,p_K$ represent the proportion of the number of subjects studied taken at time $t_1,\ldots,t_K$, respectively. For a total sample of size $n$, $n_i = np_i$ is the number of subjects to be studied at $t_i$. For the purpose of obtaining the most precise estimators of the entire model parameters, one needs to identify a design measure whose related Fisher information matrix is nonsingular and whose determinant is maximized in the class of competing designs. This is because $M^{-1}(\xi,\vec{\beta})$ is proportional to the asymptotic variance and covariance matrix of the MLE of the model parameters. An LD design for a given set of parameters has the maximum determinant for its Fisher information matrix based on the postulated values for these parameters.

When the number of PK parameters of interest is $s \leq k$ in a $k$-parameter nonlinear model, the asymptotic generalized variance of these $s$ estimators can be expressed as $n^{-1}[M(\xi, \vec{\beta})/M_{22}(\xi, \vec{\beta})]^{-1}$, where

$$\begin{pmatrix} M_{11}(\xi, \vec{\beta}) & M_{12}(\xi, \vec{\beta}) \\ M_{21}(\xi, \vec{\beta}) & M_{22}(\xi, \vec{\beta}) \end{pmatrix}$$

is the partitioned form of $M(\xi, \vec{\beta})$ with $M_{11}(\xi, \vec{\beta})$ being the associated $s \times s$ information matrix of the $s$ parameters under the design $\xi$. Following White [29], the design $\xi^*$ in a class of competing designs, $\mathcal{D}$, is said to be an $\text{LD}_s$ design if $\det M(\xi^*, \vec{\beta})/\det M_{22}(\xi^*, \vec{\beta}) = \max_{\xi \in \mathcal{D}}[\det M(\xi, \vec{\beta})/\det M_{22}(\xi, \vec{\beta})]$. White [29] established that the design $\xi^*$ is an $\text{LD}_s$ design if and only if $\sup_{t \in \mathcal{T}} d(t, \xi^*, \vec{\beta}) = s$, where $d(t, \xi, \vec{\beta}) = \text{tr}\{I(t, \vec{\beta})M^{-1}(\xi, \vec{\beta})\} - \text{tr}\{I_{22}(t, \vec{\beta}) \times M_{22}^{-1}(\xi, \vec{\beta})\}$ when $s < k$ and $d(t, \xi, \vec{\beta}) = \text{tr}\{I(t, \vec{\beta})M^{-1}(\xi, \vec{\beta})\}$ when $s = k$ and $\mathcal{T} = [0, \infty)$ is the set of all sampling time points with the understanding that the extreme right time point is a very large practitioner-selected time point. Here, $I(t, \vec{\beta})$ is the information matrix for the design with the support point of $t$ only and it is partitioned as that of $M(\xi, \vec{\beta})$. The quantity $d(t, \xi, \vec{\beta})$ can be interpreted as the variance of the estimated response at time $t$ when $s = k$ and is the variance of the estimated response after eliminating the effects of the $k - s$ nuisance parameters when $s < k$.

Before searching for optimal $t_i$ and $p_i$, the required number of design points in an LD design support is investigated first for both theoretical and practical interests. Knowing the required number of design points in advance would significantly reduce the task of searching for an LD design.

**3. Main results.** In this section, the support size of an LD design for the Emax-PK1 model under the normality assumption is investigated. Since the corresponding induced design space, $\mathcal{C} = \{\sigma^{-1}(\frac{\partial \eta(t, \vec{\beta})}{\partial \vec{\beta}}) : t \in \mathcal{T}\}$, is a bounded subset in $\mathbb{R}^k$ and the determinant of the Fisher information matrix is a continuous function on $\mathcal{C}$, thus an LD design for this setup must exist. In what follows, by a saturated design it is meant a design in which the number of design time points is equal to the number of model parameters.

THEOREM 1. *Under model (1.1):* $E(Y_t) = \beta_0 + \frac{\beta_1 \cdot D}{\beta_2 e^{\beta_3 t} + D}$ *with* $\text{Var}(Y_t) = \sigma^2$, *where* $t \in [0, u]$, *an LD design* $\xi^*$ *is a saturated D-optimal design with time* 0 *and* $u$ *in its support.*

PROOF. Let $M(\xi^*, \vec{\beta})$ be the Fisher information matrix for $\vec{\beta} = (\beta_0, \beta_1, \beta_2, \beta_3)^T$ based on an LD design $\xi^*$ and define $f_0(t) = \text{tr}(I(t, \vec{\beta})M^{-1}(\xi^*, \vec{\beta})) + c$, where $c \in \mathbb{R}$ is arbitrary and

$$I(t, \vec{\beta}) = \sigma^{-2} \begin{pmatrix} 1 \\ D(\beta_2 e^{\beta_3 t} + D)^{-1} \\ -\beta_1 D e^{\beta_3 t} (\beta_2 e^{\beta_3 t} + D)^{-2} \\ -\beta_1 D \beta_2 t e^{\beta_3 t} (\beta_2 e^{\beta_3 t} + D)^{-2} \end{pmatrix}$$

$$\times \begin{pmatrix} 1 \\ D(\beta_2 e^{\beta_3 t} + D)^{-1} \\ -\beta_1 D e^{\beta_3 t} (\beta_2 e^{\beta_3 t} + D)^{-2} \\ -\beta_1 D \beta_2 t e^{\beta_3 t} (\beta_2 e^{\beta_3 t} + D)^{-2} \end{pmatrix}^T .$$

It can be shown that $f_0(t)$ has the same number of zeros as $f(t) = a_{40} e^{4\beta_3 t} + e^{3\beta_3 t}(a_{31}t + a_{30}) + e^{2\beta_3 t}(a_{22}t^2 + a_{21}t + a_{20}) + e^{\beta_3 t}(a_{11}t + a_{10}) + a_{00}$, where $a_{ij}$ depends on $\xi^*, D$ and $\beta_k$, $i \in \{0,1,2,3,4\}$, $j \in \{0,1,2\}$, $k \in \{1,2,3\}$. Notice that $a_{22} > 0$ and $a_{31} < 0$ since the cofactor $\text{Cof}_{14}$ in $M^{-1}$ is positive (Appendix). Also, by the Remark in the Appendix, it can be shown that $a_{11} > 0$. Now in order to prove that $f(t)$ has at most six zeros, the properties of various derivatives of $f(t)$ will be explored.

By differentiating $f(t)$ with respect to $t$, one has

$$f'(t) = 4\beta_3 a_{40} e^{4\beta_3 t} + e^{3\beta_3 t}(3\beta_3 a_{31}t + b_{30})$$
$$+ e^{2\beta_3 t}(2\beta_3 a_{22}t^2 + b_{21}t + b_{20}) + e^{\beta_3 t}(\beta_3 a_{11}t + b_{10}),$$

where the $b_{ij}$'s are the corresponding constants. After setting $f'(t) = 0$ and multiplying both sides of $f'(t) = 0$ by $e^{-4\beta_3 t}$, one obtains

$$\tilde{f}'(t) = 4\beta_3 a_{40} + e^{-\beta_3 t}(3\beta_3 a_{31}t + b_{30}) + e^{-2\beta_3 t}(2\beta_3 a_{22}t^2 + b_{21}t + b_{20})$$
$$+ e^{-3\beta_3 t}(\beta_3 a_{11}t + b_{10}).$$

Notice that $f'(t)$ and $\tilde{f}'(t)$ have the same number of zeros. After differentiating $\tilde{f}'(t)$, one has

$$\tilde{f}''(t) = e^{-\beta_3 t}(-3\beta_3^2 a_{31}t + c_{30}) + e^{-2\beta_3 t}(-4\beta_3^2 a_{22}t^2 + c_{21}t + c_{20})$$
$$+ e^{-3\beta_3 t}(-3\beta_3^2 a_{11}t + c_{10}),$$

where $c_{ij}$'s are the corresponding constants. By multiplying $\tilde{f}''(t)$ by $e^{3\beta_3 t}$, one has

$$\tilde{\tilde{f}}''(t) = e^{2\beta_3 t}(-3\beta_3^2 a_{31}t + c_{30})$$
$$+ e^{\beta_3 t}(-4\beta_3^2 a_{22}t^2 + c_{21}t + c_{20}) + (-3\beta_3^2 a_{11}t + c_{10}).$$

Differentiating $\tilde{\tilde{f}}''(t)$ yields

$$\tilde{\tilde{f}}'''(t) = e^{2\beta_3 t}(-6\beta_3^3 a_{31}t + d_{30}) + e^{\beta_3 t}(-4\beta_3^3 a_{22}t^2 + d_{21}t + d_{20}) - 3\beta_3^2 a_{11},$$

where $d_{ij}$'s are the corresponding constants. Next, it will be shown that the function

$$g(t) = e^{2\beta_3 t}(-6\beta_3^3 a_{31} t + d_{30}) + e^{\beta_3 t}(-4\beta_3^3 a_{22} t^2 + d_{21} t + d_{20})$$

has at most three stationary points. By multiplying both sides of $g'(t) = 0$ by $e^{-2\beta_3 t}$, one has

$$\tilde{g}'(t) = -12\beta_3^4 a_{31} t + e_{30} + e^{-\beta_3 t}(-4\beta_3^4 a_{22} t^2 + e_{21} t + e_{20}) = 0,$$

where $e_{ij}$'s are the corresponding constants. Setting $\tilde{g}''(t) = 0$, one has

(3.1) $$e^{-\beta_3 t}(4\beta_3^5 a_{22} t^2 + f_{21} t + f_{23}) = 12\beta_3^4 a_{31},$$

where $f_{ij}$'s are the corresponding constants. Since the left-hand side of (3.1) has at most two stationary points and it approaches 0 above the $t$ axis as $t$ goes to $\infty$, the most right monotone interval of the left-hand side will not intersect the horizontal line $y = 12\beta_3^4 a_{31}$, which is below the $t$ axis. Therefore, (3.1) has at most two roots.

Now, since (3.1) has at most two roots, $\tilde{g}'(t)$ will have at most three zeros. This implies that $g(t)$ has at most three stationary points. Since $g(t)$ goes to 0 below the $t$ axis as $t$ goes to $-\infty$ and $3\beta_3^2 a_{11} > 0$, the most left monotone interval will not intersect the horizontal line $y = 3\beta_3^2 a_{11}$, which is above the $t$ axis. Consequently, $\tilde{\tilde{f}}'''(t)$ has at most three zeros. This implies that $\tilde{f}''(t)$ has at most four zeros and $f'(t)$ has at most five zeros. As a result, $f(t)$ has at most six zeros.

Since $f(t)$ has at most six zeros for all $c's$ and by the equivalence theorem the interior points of an LD design must be locally maximum points of $\text{tr}(I(t, \vec{\beta}) M^{-1}(\xi^*, \vec{\beta}))$, therefore there are at most two optimal design points on $(0, \infty)$. Otherwise, $f(t)$ has more than six zeros for some $c$. Since the induced design point at time 0 is not proportional to that at time $u > 0$, the existence of an LD design based on model (1.1) forces the two boundary points to be in the optimal design support. □

Theorem 1 demonstrates that an LD design for model (1.1) is of the form $\xi^* = \{(0, 1/4), (t_2, 1/4), (t_3, 1/4), (u, 1/4)\}$ over $t \in [0, u]$. Consequently, to search for an LD design for model (1.1), one only needs to find out $t_2$ and $t_3$.

THEOREM 2. *An LD design for the Emax-PK2 model has minimum support size with time point 0 in its support if $d(t, \xi, \vec{\beta})$ has precisely four locally maximum time points when $t \in (0, \infty)$.*

PROOF. By Carathéodory's theorem [25], for this five-parameter nonlinear model, an LD design must have at least five time points in its support.

By the equivalence theorem, LD design time points must be necessarily the locally maximum points of $d(t, \xi, \vec{\beta})$ when $t \in (0, \infty)$. Since $d(t, \xi, \vec{\beta})$ has four locally maximum time points and the induced design time points at $t = 0$ and $t = \infty$ are the same, the existence of an LD design forces time point 0 and the four locally maximum points to be in the design support. $\square$

Theorem 2 provides a sufficient condition for an LD design for the Emax-PK2 model to be minimally supported when the original design space is $[0, \infty)$. In general, if the required number of LD sampling time points is unknown for a $k$-parameter nonlinear model, the practitioner could search for the best $k$-point uniform LD design first. Then the V-algorithm by Fedorov [6] can be applied to search for an LD design with the $k$-point uniform LD design as the initial design.

When $\beta_0$ is a nuisance parameter, it is found that an LD design is also an $LD_{k-1}$ design for a $k$-parameter nonlinear model of the form $y_{tj} = \beta_0 + \tilde{\eta}(t, \vec{\beta}) + \varepsilon_{tj}$, where $\varepsilon_{tj}$ are i.i.d. $N(0, \sigma^2)$ and $\partial \tilde{\eta}(t, \vec{\beta})/\partial \beta_i$ is not free of $\beta_i$ for $i = 1, \ldots, k-1$. For a model of this form, the following result is obtained.

THEOREM 3. *An LD design for the $k$-parameter nonlinear model is also an $LD_{k-1}$ design for parameters $\beta_1, \ldots, \beta_{k-1}$.*

The proof is based on the fact that both $I_{22}$ in $I(t, \vec{\beta})$ and $M_{22}^{-1}$ in $M(\xi^*, \vec{\beta})$ are equal to 1. Theorem 3 shows that the $LD_{k-1}$ design is globally optimized for $\beta_0$ although it is only locally optimal for the other nonlinear parameters. Theorem 3 is applicable to both Emax-PK models discussed in this paper. Examples of $LD_{k-1}$ designs originated from LD designs for both Emax-PK1 and Emax-PK2 models are given in Section 4.

**4. Illustrated examples.** Maximizing the determinant of the Fisher information matrix under nonlinear models requires the foreknowledge of the values of model parameters. In practice, a good guess can be obtained from a pilot experiment. For example, suppose based on the prior information from the pilot experiment, the minimum residual effect is equal to 0.5, the maximum drug effect is equal to 10, the $ED_{50}$ is equal to 1 mg/kg and the total elimination rate is equal to 0.1 hour$^{-1}$. The search for an LD design generally contains two steps: (1) finding a best $k$-point uniform LD design; (2) using the V-algorithm (Fedorov [6]) to find an LD design given the initial design as that found in step (1). For the Emax-PK2 model, a search as described was performed. However, for the Emax-PK1 model, the search was stopped at step (1) since Theorem 1 shows that an LD design for this model is a saturated LD design.

For the Emax-PK1 model with postulated $\vec{\beta} = (0.5, 10, 1, 0.1)$ and an initial dose of 5 mg/kg, a four-point LD design is found to be uniform at times 0, 13.48, 31.62 and 160.00 hours. Figure 1 illustrates some essential characteristics of $d(t, \xi, \vec{\beta})$ for this design.

If the maximum sampling time is 72 or 24 hours, then the LD design can be found to be uniform at times 0, 13.26, 31.05 and 72 hours or at times 0, 7.97, 17.83 and 24 hours, respectively. The $d(t, \xi, \vec{\beta})$ for these two designs are illustrated in Figures 2 and 3.

Although the upper bound of the support size for an LD design based on the Emax-PK2 model is 16 by Carathéodory's theorem, one can still follow the two-step search described above. For the Emax-PK2 model with postulated $\vec{\beta} = (0.5, 10, 0.5, 0.1, 1)$, an LD design is found to be uniform at times 0, 0.275, 2.999, 14.75 and 32.7 hours. The associated $d(t,\xi,\vec{\beta})$ for this design is illustrated in Figure 4.

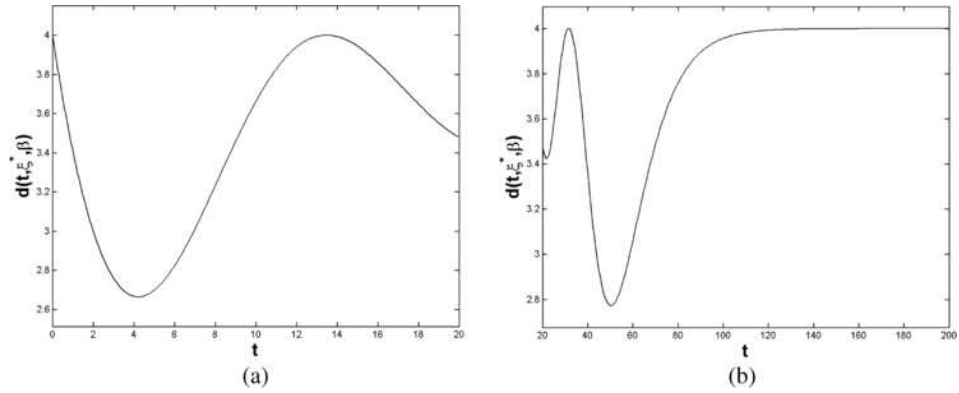

FIG. 1. *Plots of* $d(t, \xi^*, \vec{\beta})$ *with* (a) $t \in [0, 20]$ *and* (b) $t \in [20, 200]$, *respectively. The design space is* $[0, \infty)$ *and the optimal sampling times are found at* $0, 13.481506, 31.624511, 160.00199$. *The initial IV dose is 5* mg/kg.

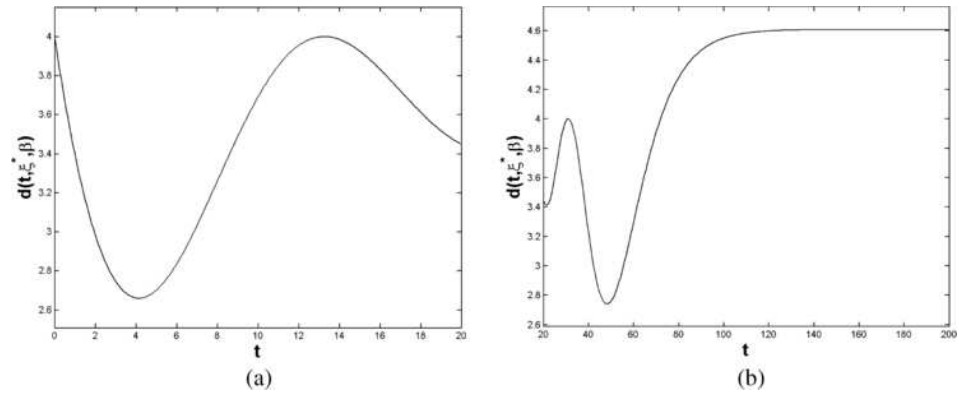

FIG. 2. *Plots of* $d(t, \xi^*, \vec{\beta})$ *with* (a) $t \in [0, 20]$ *and* (b) $t \in [20, 200]$, *respectively. The design space is* $[0, 72]$ *hours and the optimal sampling times are found to be at* $0, 13.263029, 31.050686$ *and* $72$ *hours.*

**5. Robust designs.** Applying an LD design in practice may be criticized for its local optimality. If the nominal values of the model parameters are not close to the true values, a more desirable design would be a robust design, which would lead to a better parameter estimation than an LD design while maintaining high efficiency. The relative efficiency of a design $\xi$ compared to an LD design $\xi^*$ for the postulated $\vec{\beta}$ is defined as $\det(\xi,\vec{\beta})/\det(\xi^*,\vec{\beta})$.

A robust index defined as $|(\partial(\det M(\xi,\vec{\beta}))/\partial\beta_i)^{-1}|$ for parameter $\beta_i$, $i = 0, 1, \ldots$ is introduced to compare the robustness of LD designs based on different nominal values of the model parameters. For a linear model, this robust index is $\infty$, indicating that an LD design is globally optimal for all values of the model parameters. However, for a nonlinear model, it measures the inverse of the changing rate of the determinant in the neighborhood of the postulated $\vec{\beta}$. Therefore, it is called the locally robust index (LRI). The larger the value of the LRI, the more locally robust the design.

One class of robust designs is the equally spaced uniform LD (ESULD) designs. An advantage of this class of designs is that it can be implemented very easily. It is robust for the estimation of parameters in the models that were described in Hedayat, Yan and Pezzuto [11, 12] and Hedayat, Zhong and Nie [10]. The class of ESULD designs is examined here for both Emax-PK1 and Emax-PK2 models. Table 1 shows that an ESULD design retains its robustness but loses its efficiency as the size of the design support increases. For practical applications, the five- or the six-point ESULD designs are recommended.

Since the efficiency of the ESULD designs is very low for the Emax-PK2 model, another class of robust designs is constructed based on an LD design.

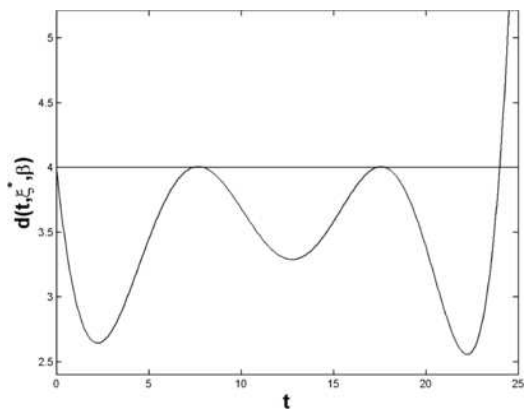

Fig. 3. *Plot of $d(t,\xi^*,\vec{\beta})$ when the design space is $[0, 24]$ hours based on the Emax-PK1 model.*

TABLE 1
*ESULD designs for the model with the initial dose = 5 mg/kg and the nominal values $\beta_1 = 10$, $\beta_2 = 1$ mg/kg, $\beta_3 = 0.1$ h$^{-1}$*

| Support size | Support of $\xi$ | Efficiency of $\xi$ | LRI for $\beta_1$ | LRI for $\beta_2$ | LRI for $\beta_3$ |
|---|---|---|---|---|---|
| 4 | $0, 13.48, 31.62, 160.00$ | 1 | 0.95901654 | 0.10344687 | 0.019187232 |
| 5 | $0, 16.176 \times i, i = 1, 2, 3, 4$ | 0.476373989 | 2.0131589 | 0.19911378 | 0.040256987 |
| 6 | $0, 14.736 \times i, i = 1, 2, 3, 4, 5$ | 0.481038755 | 1.9936368 | 0.20439842 | 0.039876038 |
| 7 | $0, 13 \times i, i = 1, 2, \ldots, 6$ | 0.440686613 | 2.1761871 | 0.22903063 | 0.043844867 |
| 8 | $0, 11.094 \times i, i = 1, 2, \ldots, 7$ | 0.403119466 | 2.3789884 | 0.25133456 | 0.047580261 |
| 9 | $0, 9.647 \times i, i = 1, 2, \ldots, 8$ | 0.376132417 | 2.549678 | 0.26807298 | 0.050989122 |
| 10 | $0, 8.558 \times i, i = 1, 2, \ldots, 9$ | 0.355042847 | 2.701129 | 0.28268401 | 0.054018207 |
| 11 | $0, 7.686 \times i, i = 1, 2, \ldots, 10$ | 0.3379349 | 2.8378736 | 0.29572619 | 0.056762451 |
| 12 | $0, 7 \times i, i = 1, 2, \ldots, 11$ | 0.323818285 | 2.9615886 | 0.30723535 | 0.058653795 |

When $s = 4$, the design is the LD design.

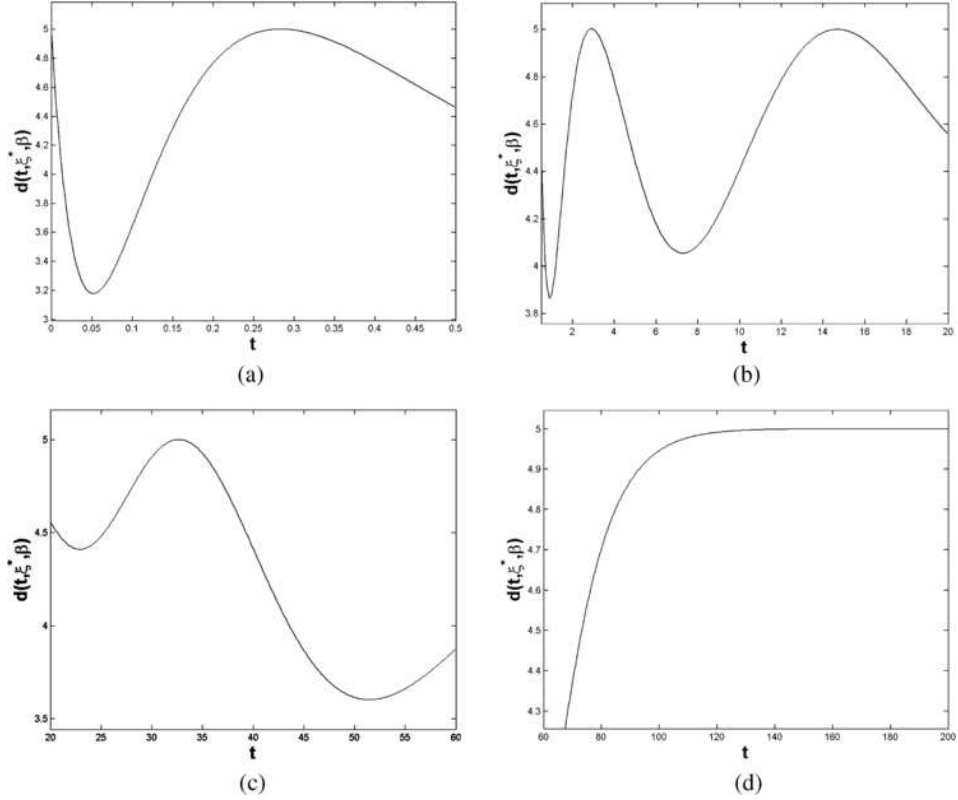

Fig. 4. *Plots of $d(t,\xi^*,\vec{\beta})$ with* (a) $t \in [0,0.5]$, (b) $t \in [0.5,20]$, (c) $t \in [20,60]$ *and* (d) $t \in [60,200]$, *respectively. The numerical solutions of the local maximums are* 0, 0.275, 2.999, 14.75, 32.695. *The initial dose is* 5 mg/kg.

The support of such a robust design includes all the design time points of an LD design as well as the design time points in the form of $t^* + r$ or $t^* - r$, where $t^*$ is a design time point of an LD design and $r$ is a fixed number. For convenience, this class of designs is referred to as equal-step expanded uniform LD (ESEULD) designs. For example, if an LD design is

$$\xi^* = \begin{pmatrix} t_1 & t_2 & t_3 \\ 1/3 & 1/3 & 1/3 \end{pmatrix},$$

then a four-point ESEULD design would be

$$\xi^{**} = \begin{pmatrix} t_1 & t_1+r & t_2 & t_3 \\ 1/4 & 1/4 & 1/4 & 1/4 \end{pmatrix};$$

a five-point ESEULD design would be

$$\xi^{**} = \begin{pmatrix} t_1 & t_1+r & t_2 & t_2+r & t_3 \\ 1/5 & 1/5 & 1/5 & 1/5 & 1/5 \end{pmatrix};$$

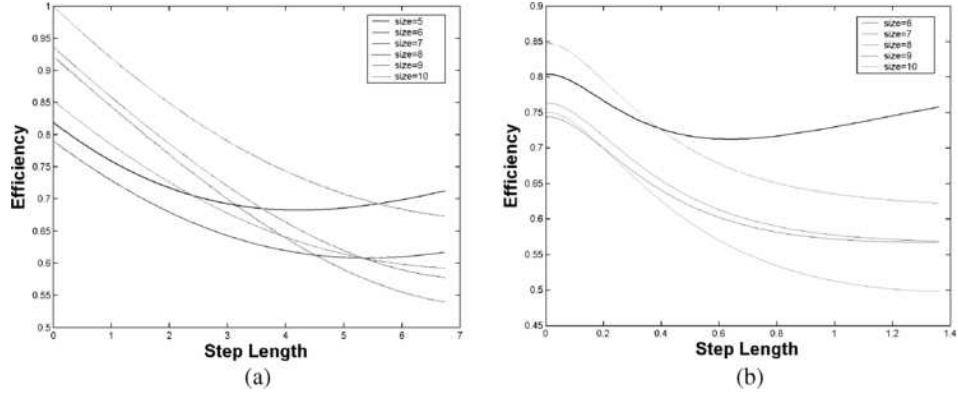

Fig. 5. *The efficiency of a ESEULD design versus step length r based on* (a) *Emax-PK1 model and* (b) *Emax-PK2 model.*

a six-point ESEULD design would be

$$\xi^{**} = \begin{pmatrix} t_1 & t_1+r & t_2 & t_2+r & t_3-r & t_3 \\ 1/6 & 1/6 & 1/6 & 1/6 & 1/6 & 1/6 \end{pmatrix}$$

and a seven-point ESEULD design would be

$$\xi^{**} = \begin{pmatrix} t_1 & t_1+r & t_2-r & t_2 & t_2+r & t_3-r & t_3 \\ 1/7 & 1/7 & 1/7 & 1/7 & 1/7 & 1/7 & 1/7 \end{pmatrix}.$$

Since the relative efficiency of the design $\xi^{**}$ relative to an LD design $\xi^*$, $\det M(\xi^{**}, \vec{\beta})/\det M(\xi^*, \vec{\beta})$, is a function of $r$ given $\xi^*$ and $\vec{\beta}$, it is impossible to give an explicit form of $r$ as a function of efficiency. However, it is possible to plot efficiency versus step length $r$ to find out the numerical relationship between $r$ and the efficiency. Figure 5(a) and 5(b) shows such plots based on the Emax-PK1 model and the Emax-PK2 model, respectively. The same postulated values of $\beta_i$'s as those in Section 4 are used for illustrations and tables throughout this section. From this figure, it is clear that the value of $r$ cannot be assigned arbitrarily for some $n$-point ESEULD designs. For example, $r = 0.95$ does not exist for a five-point ESEULD design based on the Emax-PK1 model. In this figure, $r = 1$ hour and $r = 0.2$ hour are chosen for robustness study based on the Emax-PK1 model and the Emax-PK2 model, respectively. The corresponding design time points are listed in Tables 2 and 3. The LRI's are calculated and listed in Tables 4 and 5. From these results, it appears that the higher the relative efficiency of a robust LD design, the less the robustness of the design.

**6. Conclusion and discussion.** This paper introduced a class of models by blending a PD Emax model and a PK compartmental model and studied some important features of LD, ESULD and ESEULD designs for the Emax-PK1 and the Emax-PK2 models in this class.

For Emax-PK1 model, an LD design is a saturated four-point uniform LD design with the two boundary time points of the design space in its support. Both time 0 and the upper bound of the design space are the informative time points here. This can be observed directly from the model. As $t$ approaches $\infty$, the Emax-PK1 model is reduced to $Y_{tj} = \beta_0 + \varepsilon_{tj}$. Therefore, the upper bound of the design space is an informative time point for parameter $\beta_0$. When $t = 0$, the model is reduced to $Y_{tj} = \beta_0 + \frac{\beta_1 D}{\beta_2 + D} + \varepsilon_{tj}$. Consequently, time 0 is an informative time point for $\beta_0$ as well as the ratio of $\beta_1/\beta_2$.

TABLE 2
*ESEULD designs based on the Emax-PK1 model with $r = 1$ hour*

| Support size | ESEULD design point |
|---|---|
| 4 | $0, 13.48, 31.62, 160.00$ |
| 5 | $0, 1, 13.48, 31.62, 160.00$ |
| 6 | $0, 1, 13.48, 14.48, 31.62, 160.00$ |
| 7 | $0, 1, 13.48, 14.48, 31.62, 32.62, 160.00$ |
| 8 | $0, 1, 13.48, 14.48, 31.62, 32.62, 159, 160.00$ |
| 9 | $0, 1, 12.48, 13.48, 14.48, 31.62, 32.62, 159, 160.00$ |
| 10 | $0, 1, 12.48, 13.48, 14.48, 30.62, 31.62, 32.62, 159, 160.00$ |

TABLE 3
*ESEULD designs based on the Emax-PK2 model with $r = 0.2$ hour*

| Support size | ESEULD design points |
|---|---|
| 5 | $0, 0.275, 2.999, 14.75, 32.695$ |
| 6 | $0, 0.275, 0.475, 2.999, 14.75, 32.695$ |
| 7 | $0, 0.275, 0.475, 2.999, 3.199, 14.75, 32.695$ |
| 8 | $0, 0.275, 0.475, 2.999, 3.199, 14.75, 14.95, 32.695$ |
| 9 | $0, 0.275, 0.475, 2.999, 3.199, 14.75, 14.95, 32.495, 32.695$ |
| 10 | $0, 0.275, 0.475, 2.799, 2.999, 3.199, 14.75, 14.95, 32.495, 32.695$ |

TABLE 4
*Efficiency and robustness of ESEULD designs for the Emax-PK1 model*

| ESEULD design | Efficiency of $\xi$ | LRI: $\beta_1$ | LRI: $\beta_2$ | LRI: $\beta_3$ |
|---|---|---|---|---|
| 4-point | 1 | 0.95901654 | 0.10344687 | 0.019187232 |
| 5-point | 0.7588 | 1.2638471 | 0.135847 | 0.025576467 |
| 6-point | 0.7284 | 1.3165725 | 0.13967382 | 0.025674003 |
| 7-point | 0.7846 | 1.2223363 | 0.12857595 | 0.021936078 |
| 8-point | 0.9193 | 1.0431845 | 0.10970443 | 0.018685843 |
| 9-point | 0.8583 | 1.1172873 | 0.1190755 | 0.020901414 |
| 10-point | 0.8437 | 1.1366831 | 0.12194774 | 0.022945661 |

TABLE 5
*Efficiency and robustness of ESEULD designs for the Emax-PK2 model*

| ESEULD design | Efficiency of $\xi$ | LRI: $\beta_1$ | LRI: $\beta_2$ | LRI: $\beta_3$ | LRI: $\beta_4$ |
|---|---|---|---|---|---|
| 5-point | 1 | 0.3067854 | 0.046489413 | 0.0090373459 | 0.043066663 |
| 6-point | 0.7667 | 0.40014335 | 0.055507767 | 0.011777852 | 0.058844405 |
| 7-point | 0.6995 | 0.43857707 | 0.059457229 | 0.013001578 | 0.064883822 |
| 8-point | 0.7168 | 0.42799782 | 0.05768681 | 0.012643369 | 0.063180006 |
| 9-point | 0.7952 | 0.38580638 | 0.052011575 | 0.011571927 | 0.057033582 |
| 10-point | 0.6995 | 0.4385802 | 0.059583199 | 0.013097578 | 0.064855939 |

A sufficient condition for the Emax-PK2 model to be minimally supported is given. Time 0 is an informative time point here. This also can be explained from the Emax-PK2 model directly. As time goes to 0, the Emax-PK2 model is reduced to $Y_{tj} = \beta_0 + \varepsilon_{tj}$.

When $\beta_0$ is considered as a nuisance parameter, an LD design and an $LD_{k-1}$ design based on any of the Emax-PK models are equivalent. The corresponding LD design is globally optimal for the linear parameter $\beta_0$ and locally optimal for the other nonlinear parameters.

Future research for the Emax-PK models could involve random effects for certain PK parameters since these parameters likely differ between subjects. The reader is referred to Mentre, Mallet and Baccar [20], Palmer and Muller [21], Han and Chaloner [9] for recent results in this area.

## APPENDIX: PROOFS

**$\text{Cof}_{14}$ of $M(\xi^*, \vec{\beta})^{-1}$ is positive and $\text{Cof}_{24}$ of $M(\xi^*, \vec{\beta})^{-1}$ is negative.** Since the Fisher information matrix of $\vec{\beta} = (\beta_0, \beta_1, \beta_2, \beta_3)^T$ at sampling time $t_i$ for an LD design $\xi^* = \begin{pmatrix} t_1 & t_2 & \cdots & t_s \\ p_1 & p_2 & \cdots & p_s \end{pmatrix}$ based on model (1.1) or the Emax-PK1 model is

$$I(t_i, \vec{\beta}) = \sigma^{-2} p_i \begin{pmatrix} 1 \\ D(\beta_2 e^{\beta_3 t_i} + D)^{-1} \\ -\beta_1 D e^{\beta_3 t_i}(\beta_2 e^{\beta_3 t_i} + D)^{-2} \\ -\beta_1 D \beta_2 t_i e^{\beta_3 t_i}(\beta_2 e^{\beta_3 t_i} + D)^{-2} \end{pmatrix} \times \begin{pmatrix} 1 \\ D(\beta_2 e^{\beta_3 t_i} + D)^{-1} \\ -\beta_1 D e^{\beta_3 t_i}(\beta_2 e^{\beta_3 t_i} + D)^{-2} \\ -\beta_1 D \beta_2 t_i e^{\beta_3 t_i}(\beta_2 e^{\beta_3 t_i} + D)^{-2} \end{pmatrix}^T,$$

the $\text{Cof}_{14}$ has the same sign as

$$-\det m_{14} = -\det \begin{pmatrix} \sum_{i=1}^{s} p_i \frac{D}{(\beta_2 e^{\beta_3 t_i} + D)} & \sum_{i=1}^{s} p_i \frac{D^2}{(\beta_2 e^{\beta_3 t_i} + D)^2} & \sum_{i=1}^{s} p_i \frac{-D^2 \beta_1 e^{\beta_3 t_i}}{(\beta_2 e^{\beta_3 t_i} + D)^3} \\ \sum_{i=1}^{s} p_i \frac{-D\beta_1 e^{\beta_3 t_i}}{(\beta_2 e^{\beta_3 t_i} + D)^2} & \sum_{i=1}^{s} p_i \frac{-D^2 \beta_1 e^{\beta_3 t_i}}{(\beta_2 e^{\beta_3 t_i} + D)^3} & \sum_{i=1}^{s} p_i \frac{D^2 \beta_1^2 e^{2\beta_3 t_i}}{(\beta_2 e^{\beta_3 t_i} + D)^4} \\ \sum_{i=1}^{s} p_i \frac{-D\beta_1 \beta_2 t_i e^{\beta_3 t_i}}{(\beta_2 e^{\beta_3 t_i} + D)^2} & \sum_{i=1}^{s} p_i \frac{-D^2 \beta_1 \beta_2 t_i e^{\beta_3 t_i}}{(\beta_2 e^{\beta_3 t_i} + D)^3} & \sum_{i=1}^{s} p_i \frac{D^2 \beta_1^2 \beta_2 t_i e^{2\beta_3 t_i}}{(\beta_2 e^{\beta_3 t_i} + D)^4} \end{pmatrix}.$$

Let

$$g(t_i, t_j, t_k) = \det \begin{pmatrix} \frac{D}{(\beta_2 e^{\beta_3 t_i} + D)} & \frac{D^2}{(\beta_2 e^{\beta_3 t_j} + D)^2} & \frac{-D^2 \beta_1 e^{\beta_3 t_k}}{(\beta_2 e^{\beta_3 t_k} + D)^3} \\ \frac{-D\beta_1 e^{\beta_3 t_i}}{(\beta_2 e^{\beta_3 t_i} + D)^2} & \frac{-D^2 \beta_1 e^{\beta_3 t_j}}{(\beta_2 e^{\beta_3 t_j} + D)^3} & \frac{D^2 \beta_1^2 e^{2\beta_3 t_k}}{(\beta_2 e^{\beta_3 t_k} + D)^4} \\ \frac{-D\beta_1 \beta_2 t_i e^{\beta_3 t_i}}{(\beta_2 e^{\beta_3 t_i} + D)^2} & \frac{-D^2 \beta_1 \beta_2 t_j e^{\beta_3 t_j}}{(\beta_2 e^{\beta_3 t_j} + D)^3} & \frac{D^2 \beta_1^2 \beta_2 t_k e^{2\beta_3 t_k}}{(\beta_2 e^{\beta_3 t_k} + D)^4} \end{pmatrix}$$

$\forall i < j < k \leq s$, then the determinant of $m_{14}$ is $\det m_{14} = \sum_{i<j<k\leq s} \sum_{\text{permutations of } i,j,k} p_i p_j p_k g(t_i, t_j, t_k)$.

Now without loss of generality, let $i=1, j=2, k=3$; then it can be shown that

$$\sum_{\text{permutations of } 1,2,3} g(t_1, t_2, t_3) = -\frac{D^7 \beta_1^3 \beta_2^3}{(\beta_2 e^{\beta_3 t_1} + D)^4 (\beta_2 e^{\beta_3 t_2} + D)^4 (\beta_2 e^{\beta_3 t_3} + D)^4}$$

$$\times \begin{vmatrix} 1 & e^{\beta_3 t_1} & t_1 e^{\beta_3 t_1} \\ 1 & e^{\beta_3 t_2} & t_2 e^{\beta_3 t_2} \\ 1 & e^{\beta_3 t_3} & t_3 e^{\beta_3 t_3} \end{vmatrix} \begin{vmatrix} 1 & e^{\beta_3 t_1} & e^{2\beta_3 t_1} \\ 1 & e^{\beta_3 t_2} & e^{2\beta_3 t_2} \\ 1 & e^{\beta_3 t_3} & e^{2\beta_3 t_3} \end{vmatrix},$$

since

$$g(t_1, t_2, t_3) = -\frac{D^6 \beta_1^3 \beta_2}{(\beta_2 e^{\beta_3 t_1} + D)^4 (\beta_2 e^{\beta_3 t_2} + D)^4 (\beta_2 e^{\beta_3 t_3} + D)^4} \begin{vmatrix} 1 & e^{\beta_3 t_1} & t_1 e^{\beta_3 t_1} \\ 1 & e^{\beta_3 t_2} & t_2 e^{\beta_3 t_2} \\ 1 & e^{\beta_3 t_3} & t_3 e^{\beta_3 t_3} \end{vmatrix}$$

$$\times (\beta_2 e^{\beta_3 t_1} + D)^2 (\beta_2 e^{\beta_3 t_2} + D) e^{\beta_3 t_3}.$$

Now since both $\{1, e^{\beta_3 t}, te^{\beta_3 t}\}$ and $\{1, e^{\beta_3 t}, e^{2\beta_3 t}\}$ are Chebyshev systems (see Karlin and Studden [14]) and their determinants are positive, $\sum_{\text{permutations of } 1,2,3} g(t_1, t_2, t_3) < 0$. Therefore, $\det m_{14} = \sum_{i<j<k\leq s} \sum_{\text{permutations of } i,j,k} p_i p_j p_k g(t_i, t_j, t_k) < 0$. This implies that the $\text{Cof}_{14} > 0$.

Next, $\text{Cof}_{24} < 0$ will be shown. By the same argument, it can be proved that

$$\text{Cof}_{24} = \sum_{i<j<k\leq s} p_i p_j p_k Q \begin{vmatrix} 1 & t_i & t_i e^{\beta_3 t_i} \\ 1 & t_j & t_j e^{\beta_3 t_j} \\ 1 & t_k & t_k e^{\beta_3 t_k} \end{vmatrix} \begin{vmatrix} 1 & e^{\beta_3 t_i} & e^{2\beta_3 t_i} \\ 1 & e^{\beta_3 t_j} & e^{2\beta_3 t_j} \\ 1 & e^{\beta_3 t_k} & e^{2\beta_3 t_k} \end{vmatrix} - \text{Cof}_{14},$$

where

$$Q = -\frac{D^5 \beta_1^3 \beta_2^5 e^{\beta_3(t_i+t_j+t_k)}}{(\beta_2 e^{\beta_3 t_i} + D)^4 (\beta_2 e^{\beta_3 t_j} + D)^4 (\beta_2 e^{\beta_3 t_k} + D)^4}.$$

Since $\{1, t, e^{\beta_3 t}\}$ is a Chebyshev system (see Karlin and Studden [14]) and $\text{Cof}_{14} > 0$, this yields $\text{Cof}_{24} < 0$.

REMARK.

$$-\text{Cof}_{14} - \text{Cof}_{24} = -\sum_{i<j<k\leq s} p_i p_j p_k Q \begin{vmatrix} 1 & t_i & t_i e^{\beta_3 t_i} \\ 1 & t_j & t_j e^{\beta_3 t_j} \\ 1 & t_k & t_k e^{\beta_3 t_k} \end{vmatrix} \begin{vmatrix} 1 & e^{\beta_3 t_i} & e^{2\beta_3 t_i} \\ 1 & e^{\beta_3 t_j} & e^{2\beta_3 t_j} \\ 1 & e^{\beta_3 t_k} & e^{2\beta_3 t_k} \end{vmatrix},$$

which is positive.

**Acknowledgments.** The authors are grateful for the many valuable comments and suggestions provided by the Associate Editor and the three referees who critically evaluated the original submission. The authors also thank Professor Bikas K. Sinha for his valuable inputs while the authors were exploring this line of research.

Department of Mathematics,
   Statistics, and Computer Sciences
University of Illinois
322 Science and Engineering Offices (M/C 249)
850 S. Morgan Street
Chicago, Illinois 60607-7045
USA
E-mail: hedayat@uic.edu